# Splitting of a nondegenerate phonon mode in PrFe$_3$(BO$_3$)$_4$: 4f crystal-field level between TO and LO phonon frequencies

K. N. Boldyrev[1], T. N. Stanislavchuk[2], A. A. Sirenko[2], L. N. Bezmaternykh[3], and M. N. Popova[1]

[1]*Institute of Spectroscopy, RAS, Moscow, Troitsk, 142190, Russia*
[2]*Department of Physics, New Jersey Institute of Technology, Newark, New Jersey 07102, USA*
[3]*Kirenskiy Institute of Physics, Siberian Branch of RAS, Krasnoyarsk, 660036, Russia*

A new effect originating from the crystal-field-excitation – phonon coupling was observed (in the far infrared spectra of a multiferroic PrFe$_3$(BO$_3$)$_4$). The reststrahlen band corresponding to the A$_2$ symmetry *nondegenerate* phonon mode near 50 cm$^{-1}$ (1.5 THz) splits into two bands at about 100 K, well above $T_N$ = 32 K. These bands shift and narrow progressively with further lowering the temperature, demonstrating pronounced peculiarities at T$_N$. The observed effects were explained by an interaction of the A$_2$ phonon mode with the 4f crystal-field electronic excitation of Pr$^{3+}$ whose frequency falls into the TO – LO frequency region of the phonon mode. Inversion of the TO and LO frequencies for the electronic excitation and a formation of coupled electron-phonon modes are discussed. Fitting of the TO frequency vs temperature experimental plots by theoretical curves revealed the value 14.6 cm$^{-1}$ for the electron-phonon coupling constant.

The interaction between crystal-field electronic excitations and lattice phonons in concentrated transition-metal (i.e., elements with partially filled *d*- or *f*-electronic shells) compounds leads to a series of interesting and important phenomena like the cooperative Jahn – Teller effect (for a recent review see Ref. [1]), magnetic phonon splitting [2-4], delocalization of the electronic states in the energy range of optical phonons and, as a consequence, the electronic Davydov splitting [3], formation of coupled electron-phonon modes accompanied by a mutual energy renormalization and appearance of new branches in the excitation spectrum [5-7]. A gap in the spectrum of elementary excitations due to repulsion of dispersion curves of the electronic and phonon systems (the so called "anticrossing" effect) was observed in a series of different experiments. In Raman scattering and infrared absorption or reflection measurements that probe the k = 0 point of the Brillouin zone, the anticrossing and transfer of transition intensities between components of the coupled electron-phonon excitations were observed via tuning of an electronic level into a resonance with a phonon, by an external magnetic field [3,4,6-8]. In neutron scattering experiments at a constant magnetic field (which determined an energy of the lowest crystal-field excitation), repulsed dispersion curves ω$_1$(k) and ω$_2$(k) were measured directly and the wave-vector dependence of the electron-phonon interaction was studied [5].

In this Letter, we report on a new effect of this type. A gap in the spectrum of excitations develops with lowering the temperature, due to a growing interaction between a nondegenerate phonon and an electronic 4f crystal-field excitation whose frequency falls into the region between TO and LO frequencies of the phonon. The effect was observed in the infrared spectra of PrFe$_3$(BO$_3$)$_4$, a compound from the family of new multiferroics with the general formula $R$Fe$_3$(BO$_3$)$_4$ ($R$ stands for a rare earth or yttrium). $R$Fe$_3$(BO$_3$)$_4$ compounds were intensively studied during recent years because of interesting physical properties and potential applications (see, e.g., Refs. [9-12]). Charge-lattice-spin coupling plays a key role in a vast variety of phases and phenomena observed in multiferroics [13]. While the spin-phonon interactions in multiferroic compounds were rather thoroughly studied, including optical studies [14,15], no data are available, as far as we know, on the interaction of crystal-field electronic excitations with optical phonons in multiferroics.

The $R$Fe$_3$(BO$_3$)$_4$ multiferroic compounds possess a huntite-type noncentrosymmetric trigonal structure that consists of helical chains of edge-sharing FeO$_6$ octahedra running along the *c*-axis of the crystal, interconnected by two kinds of BO$_3$ triangles and $R$O$_6$ distorted prisms [16]. In the case of $R$Fe$_3$(BO$_3$)$_4$ $R$ = Pr, Nd, and Sm, the structure is described by the *R32* space group at all the temperatures [17,11,10,18]. There is only one single $D_3$ symmetry position for the rare-earth (RE) ion in this space group. Crystal-field (CF) levels of a non Kramers ion (i.e., the ion with an even number of electrons, like Pr$^{3+}$) are characterized by the Γ$_1$ and Γ$_2$ nondegenerate and Γ$_3$ doubly degenerate irreducible representations of the $D_3$ point symmetry group.

The presence of two interacting magnetic subsystems (Fe and RE) results in a great variety of magnetic and magnetoelectric properties of $R$Fe$_3$(BO$_3$)$_4$ compounds, depending on a specific RE ion (see, e.g., Ref. 11(b) and references therein). In particular, PrFe$_3$(BO$_3$)$_4$ orders at $T_N$ = 32 ± 1 K into the easy-axis antiferromagnetic structure [17,11]. Magnetic and magnetoelectric properties of PrFe$_3$(BO$_3$)$_4$ are governed, mainly, by the singlet ground Γ$_2$ and the first excited Γ$_1$ (at about 48 cm$^{-1}$) states of the Pr$^{3+}$ ion [17,11,12]. Intermixing of these two lowest CF states by the exchange interaction of Pr$^{3+}$ with an ordered Fe subsystem causes a shift of the Γ$_1$ (48 cm$^{-1}$) level and gives rise to the breaking of selection rules for optical transitions below $T_N$ and appearance of forbidden spectral lines corresponding to the *f* - *f* transitions of Pr$^{3+}$ [11].

The energy of the Pr$^{3+}$ CF level 48 cm$^{-1}$ is very close to the energy ħω$_0$ of the lowest-frequency infrared-active phonon associated with motions of the $R^{3+}$ ions in $R$Fe$_3$(BO$_3$)$_4$ compounds [19], so that pronounced effects due to the electron-phonon interaction could be anticipated when probed by the terahertz (far infrared, FIR) radiation (see, e.g., Refs. [6,7]). Interaction of low-frequency lattice vibrations with CF excitations in RE containing multiferroic compounds can substantially change dispersion of original excitations and cause an onset of new bands in the excitation spectrum, which in its turn can

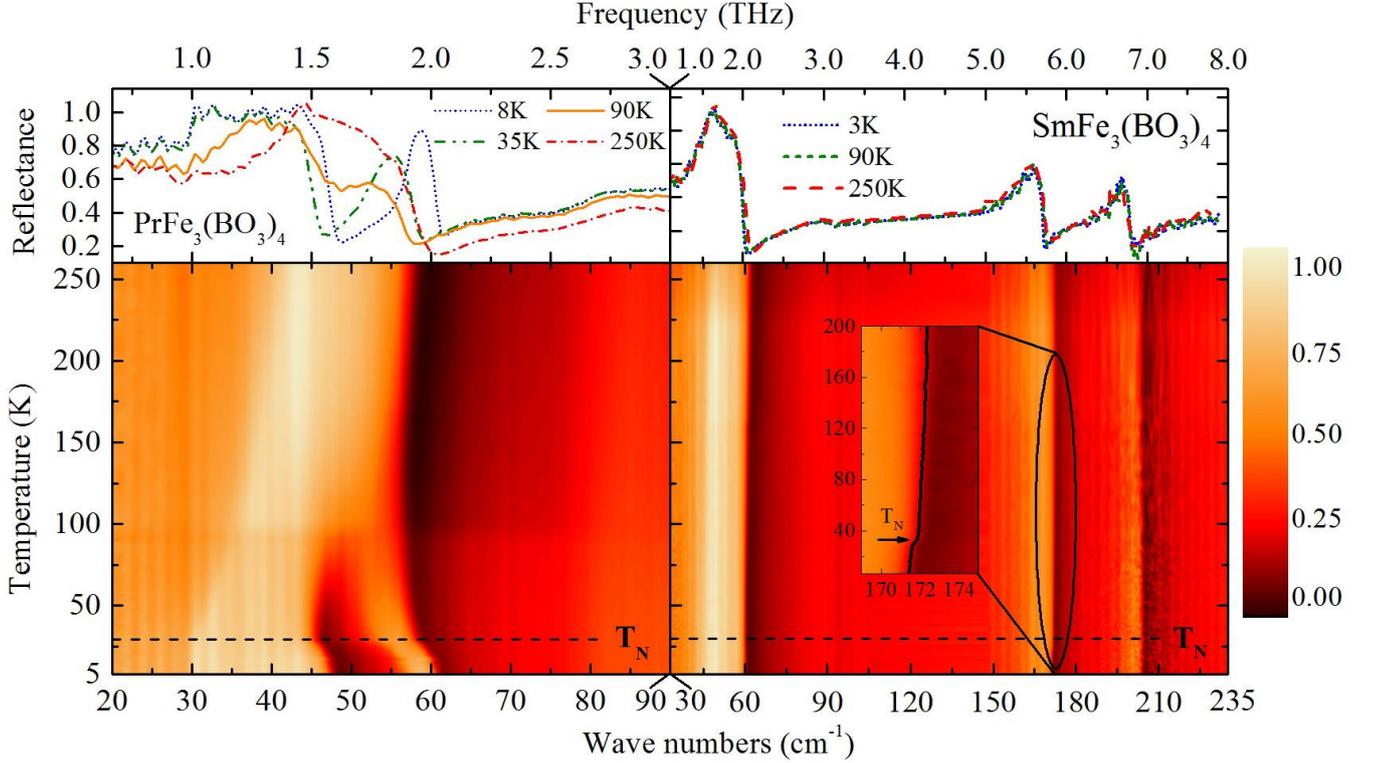

FIG. 1 (color online). The π-polarized FIR reflection spectra (upper parts) and the corresponding reflection intensity maps in the frequency-temperature axes (lower parts) for (a) $PrFe_3(BO_3)_4$ and (b) $SmFe_3(BO_3)_4$, $T_N = 32 \pm 1$ K for both compounds. A splitting of the $PrFe_3(BO_3)_4$ reststrahlen band near 50 cm$^{-1}$ below ~ 100 K is seen. Inset in (b) reveals a small kink at $T_N$ in the $\omega(T)$ dependence for the $A_2^2$ phonon mode of $SmFe_3(BO_3)_4$.

influence magnetodielectric properties of a multiferroic. In the present work, we study electron-phonon interaction in a multiferroic $PrFe_3(BO_3)_4$ single crystal by means of FIR reflection and ellipsometry measurements in the spectral region 0.6 - 3 THz (20 – 100 cm$^{-1}$), in a broad range of temperatures (2 – 300 K).

A $PrFe_3(BO_3)_4$ single crystal of good optical quality was grown in the Kirensky Institute of Physics in Krasnoyarsk from solution-melts in a $Bi_2Mo_3O_{12}$ based flux, as described in Ref. [11]. The sample with dimensions 5x5x10 mm was oriented using the crystal morphology and optical polarization methods. A Fourier spectrometer Bruker IFS 125 HR with a liquid helium bolometer (4.2 K) as a detector and a closed helium cycle cryostat Cryomech ST403 were used to register optical reflectance spectra in the region 20-100 cm$^{-1}$ (0.6 – 3 THz) at the temperatures 4 –300 K. Spectra were studied in the π ($\mathbf{k} \perp c$, $\mathbf{E} \parallel c$, $\mathbf{H} \perp c$) and σ ($\mathbf{k} \perp c$, $\mathbf{E} \perp c$, $\mathbf{H} \parallel c$) polarizations. FIR ellipsometry measurements were also performed, using self-made ellipsometer on the U4IR beamline of the National Synchrotron Light Source, Brookhaven National Laboratory, USA [20].

Far-infrared transmission and reflection spectra of crystals containing RE ions probe optical phonons from the center of the Brillouin zone (i.e., with the wave vector $k$=0), as well as electronic excitations corresponding to optical transitions between the CF levels of the ground multiplet of the RE ion. The electron-phonon interaction intermixes both types of excitations, so that instead of purely vibrational and purely electronic excitations one deals with a coupled electron-phonon modes or a quasi-phonon or quasi-electronic excitations, depending on a strength of the electron-phonon coupling. Only the excitations of the same symmetry interact, the strongest coupling takes place if the frequencies of a purely phonon mode and of a purely electronic excitation coincide.

RE borates that belong to the $R32$ ($D_7^3$) space symmetry group (like $PrFe_3(BO_3)_4$) have one formula unit in a primitive crystal cell (20 atoms), so their vibrational spectrum consists of 60 branches. 57 optical Γ point ($k$=0) phonons are characterized by irreducible representations of the crystal factor group $D_3$ as follows [10]: $\Gamma_{vibr} = 7 A_1 + 12 A_2(z) + 19 E(xy)$. The $A_2$ ($E$) phonon modes are IR active for the electric vector of radiation polarized along (perpendicular to) the $c$ crystalline axis. $E$ modes are also Raman active, as well as $A_1$ modes. While the Raman spectra of some $RFe_3(BO_3)_4$ ($R$ = Nd, Gd, Tb, Er, Y) were studied earlier [10], no data existed on IR active phonons in RE iron borates. Our FIR reflection experimental data on $PrFe_3(BO_3)_4$ at room temperature show that there are only two vibrational modes in the studied frequency range (20 – 100 cm$^{-1}$), namely, one $A_2(z)$ vibration near 45 cm$^{-1}$ and one $E(xy)$ doubly degenerate vibration at 84 cm$^{-1}$. As the point symmetry group of the RE site is the same as the crystal factor group, it follows immediately that the $A_2(z)$ phonons interact with the $\Gamma_2 - \Gamma_1$ and $\Gamma_3 - \Gamma_3$ electronic transitions between the CF states of $Pr^{3+}$ but $E(xy)$ phonons interact with the $\Gamma_1 - \Gamma_3$, $\Gamma_2 - \Gamma_3$, and $\Gamma_3 - \Gamma_3$ transitions. Only one excited CF level of $Pr^{3+}$ is present in the studied frequency range, namely, the $\Gamma_1$ level at ~ 48 cm$^{-1}$ [11], so that the $\Gamma_2 - \Gamma_1$ electronic transition, allowed as an electric dipole one in the π polarization, interacting with the $A_2(z)$ phonon could be the only candidate to observe effects of

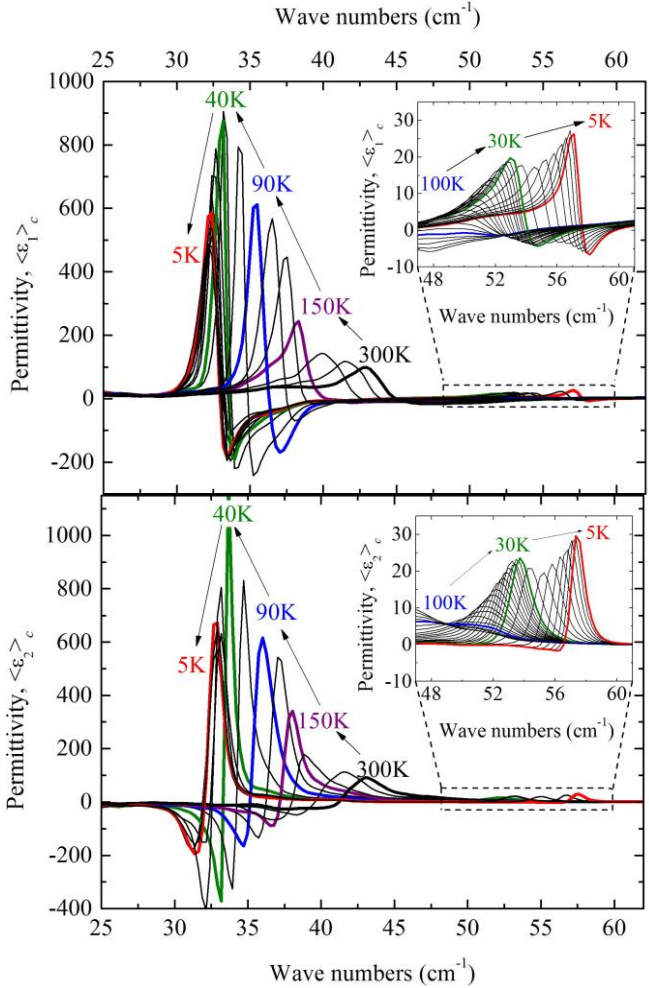

FIG. 2 (color online). The real $\langle\varepsilon_1(\omega)\rangle$ and imaginary $\langle\varepsilon_2(\omega)\rangle$ parts of the pseudo-dielectric constant of PrFe$_3$(BO$_3$)$_4$ obtained from the ellipsometry data at different temperatures. Insets show an expanded view of the emerging high-frequency branch of the spectrum.

the electron-phonon coupling in our experiments.

The upper part of Fig. 1a shows π-polarized reflection spectra of PrFe$_3$(BO$_3$)$_4$ at several selected temperatures but the lower part of the same figure displays an intensity map smoothly scanned vs temperature. At room temperature, a strong reststrahlen reflection band typical for a phonon is observed. Fitting of π-polarized reflection spectra in a broad spectral range occupied by twelve $A_2$ modes by the oscillator model according to the relations

$$R = \left|\frac{\sqrt{\varepsilon}-1}{\sqrt{\varepsilon}+1}\right|^2 \quad (1)$$

$$\varepsilon(\omega) = \varepsilon_\infty + \sum_i \frac{\Delta\varepsilon_i \omega_{0i}^2}{\omega^2 - \omega_{0i}^2 - i\omega\gamma_i}, \quad (2)$$

where $\omega_{0i}$ are the phonon TO frequencies, $\gamma_i$ are the damping constants, $\Delta\varepsilon_i$ are the oscillator strengths, $\varepsilon_\infty$ is the dielectric constant at high frequencies, has revealed the following parameters for the studied lowest frequency $A_2^1$ phonon at room temperature: $\omega_{TO} \approx \omega_{01} \equiv \omega_0 = 44.6$ cm$^{-1}$, $\omega_{LO} = 58.8$ cm$^{-1}$ [LO frequencies were found as zeros of $\varepsilon(\omega)$], $\gamma = 2.8$ cm$^{-1}$, $\Delta\varepsilon = 3.6$. These parameters are similar to the respective parameters for the isostructural SmFe$_3$(BO$_3$)$_4$, $\omega_{TO} = 48.6$ cm$^{-1}$, $\omega_{LO} = 61.5$ cm$^{-1}$, $\gamma = 2.4$ cm$^{-1}$, $\Delta\varepsilon = 2.9$. At about 100 K, a splitting of the reststrahlen band of PrFe$_3$(BO$_3$)$_4$ is observed. Components of the split band shift in the opposite directions with further lowering the temperature. A pronounced peculiarity in the behavior of both components is observed at the magnetic ordering temperature $T_N$. In contrast, for SmFe$_3$(BO$_3$)$_4$, the same phonon mode does not show any splitting and only a small kink at $T_N$ can be found in $\omega(T)$ dependences of several phonon modes (see Fig. 1b), similar to the case of EuFe$_3$(BO$_3$)$_4$ [15].

Figure 2 shows the real $\langle\varepsilon_1(\omega)\rangle$ and imaginary $\langle\varepsilon_2(\omega)\rangle$ parts of the pseudo-dielectric constant of PrFe$_3$(BO$_3$)$_4$ obtained from the ellipsometry data at different temperatures. Position of the peak in $\langle\varepsilon_2\rangle$ coincides with the TO frequency, the width is proportional to the damping constant. Figure 2 clearly demonstrates a shift and a narrowing of the quasi-phonon mode with lowering the temperature from RT to ~ 40 K and a progressive loss of its intensity below ~ 40 K. The quasi-electronic mode that appears below ~ 100 K (see also Fig. 1), evidently, gains its intensity from the quasi-phonon mode. A pronounced shift of the quasi-electronic mode to higher frequencies is observed below the temperature of an antiferromagnetic ordering $T_N$ (see Fig.1 and Inset of Fig.2b).

Using the FIR reflection and ellipsometry data, we have plotted the $\omega_{TO}(T)$ and $\omega_{LO}(T)$ dependences for the quasi-phonon $A_2^1$ and the quasi-electronic modes (see Fig. 3). The frequency vs temperature dependence for the σ-polarized $E(xy)$ phonon mode near 84 cm$^{-1}$ is also shown in Fig. 3. This mode exhibits an almost linear softening with lowering the temperature, from 84 cm$^{-1}$ at 315 K to 81 cm$^{-1}$ at 5 K. Such behavior contradicts to a hardening of force constants due to a diminishing of interatomic distances in a crystal with cooling and could be a consequence of either an interaction of the $E^1(xy)$ lattice vibration with the $\Gamma_2 - \Gamma_3$ electronic excitation corresponding to the transition from the ground state to the lowest $\Gamma_3$ CF state at 192 cm$^{-1}$ [11(b)] or an interaction of this phonon mode with the spin system, due to a modulation of Fe-O-Fe angles by a given vibration [20].

However, as it has already been mentioned before, the most pronounced effects due to the electron-phonon interaction are observed for the lowest-frequency $A_2^1$ phonon and the $\Gamma_2 - \Gamma_1$ electronic excitation with nearly the same energy. Frequencies of these coupled electron-phonon excitations in PrFe$_3$(BO$_3$)$_4$ can be found as roots of the following equation obtained on the basis of results derived using the Green's functions method and published in Ref. [22]:

$$\omega^2 - \omega_0^2 + \frac{2\omega_0\omega_{12}(n_1-n_2)|W|^2}{\omega^2 - \omega_{12}^2} = 0 \quad (4)$$

Here $\omega_0$ and $\omega_{12}$ are the frequencies (in cm$^{-1}$) of the vibrational and electronic excitations, respectively, in the absence of interaction; $n_1$ and $n_2$ are relative populations of the excited $|\Gamma_1\rangle$ and ground $|\Gamma_2\rangle$ CF states of Pr$^{3+}$, respectively; $W$ is the interaction constant between the electronic excitation $\omega_{12}$ and the Γ-point $A_2^1$ optical phonon. This constant determines a change of the RE ion's

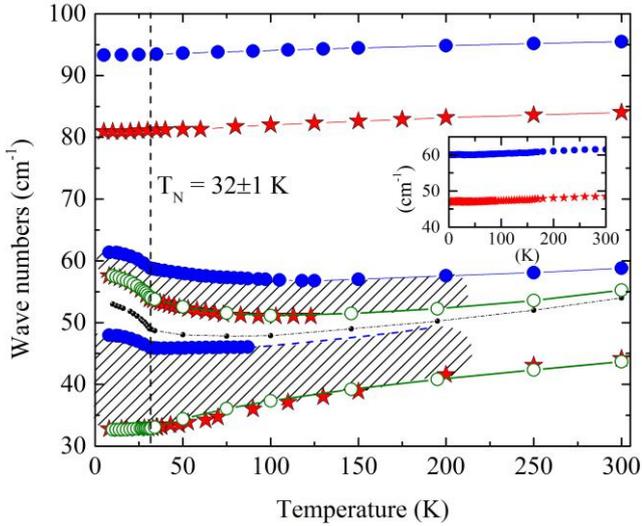

FIG. 3 (color online). Temperature dependences of the TO (red stars) and LO (blue circles) frequencies extracted from the FIR reflection and ellipsometry data for PrFe$_3$(BO$_3$)$_4$ and SmFe$_3$(BO$_3$)$_4$ (Inset). Small black balls represent the temperature-dependent position of the Pr$^{3+}$ crystal-field level found earlier from optical spectroscopy data [11]. Thin solid and dashed lines are guides for the eye. The calculated (from Eq. 6) TO frequencies of the coupled electron-phonon modes are shown by green open circles and thick lines.

energy due to a modulation of the crystal field by the $A_2^1$ lattice vibration and can be represented in the following form [7]:

$$W \equiv W_{12}(A_2^1) = \langle \Gamma_1 | \tilde{B}_3^4 O_3^4 + \tilde{B}_3^4 O_3^6 + \tilde{B}_3^4 O_{-6}^6 | \Gamma_2 \rangle = W_0 e^{i\eta} \quad (5)$$

where $O_q^p$ are the Stevens operators of the symmetry $\Gamma_2$ (or $A_2$ in the notations accepted for phonons), $\tilde{B}_q^p = \tilde{B}_q^p(A_2^1)$ are the interaction constants of the Pr$^{3+}$ ion in PrFe$_3$(BO$_3$)$_4$ with the $\Gamma$-point $A_2^1$ optical phonon, they can be calculated as derivatives of the CF parameters with respect to ionic displacements corresponding to the $A_2^1$ zone-center lattice mode [7]. Solution of Eq. (4) yields:

$$\omega_\pm^2 = \frac{\omega_0^2 + \omega_{12}^2}{2} \pm \sqrt{\frac{(\omega_{12}^2 - \omega_0^2)^2}{4} + 2\omega_0 \omega_{12}(n_2 - n_1)|W|^2} \quad (6)$$

At high temperatures, $n_1 \approx n_2$, the electron-phonon interaction vanishes, and we have pure phonon and electronic excitations with frequencies $\omega_+ = \omega_{12}$ and $\omega_- = \omega_0$, respectively.

Now, we use Eq. 6 to fit the experimental data of Fig. 3. In the case of the Boltzman distribution of populations of electronic levels, the difference of populations $n_1 - n_2$ is given by $n_1 - n_2 = th \frac{\omega_{12}(T)}{2kT}$. The function $\omega_{12}(T)$ coincides with the temperature-dependent position of the Pr$^{3+}$ crystal-field level found earlier from optical spectroscopy data [11]. The interaction constant $W$ defined by Eq. 5 may depend on temperature below $T_N$, because of the intermixing of the $|\Gamma_1\rangle$ and $|\Gamma_2\rangle$ CF states by an internal magnetic field $B_{ex}(T)$ created by ordered Fe magnetic moments. Neglecting higher electronic states, we obtain the following relation [23]:

$$W(T)^2 = W_0{}^2 [1 - 2|\alpha(T)|^2 (\cos 2\varphi + 1)/(1+4|\alpha(T)|^2)]. \quad (7)$$

Here $W_0$ is the interaction constant in the paramagnetic phase of PrFe$_3$(BO$_3$)$_4$; the quantity $\alpha = \alpha(T)$ is defined as $\alpha = V_{12}(T)/\omega_{12}(T_N)$, where the matrix element of the Zeeman interaction $V_{12}(T)$ is given by

$$V_{12}(T) = \mu_B g_0 B_{ex}(T) \langle \Gamma_1 | J_z | \Gamma_2 \rangle = V_0 e^{i\varsigma} \quad (8)$$

($\mu_B$ is the Bohr magneton, $g_0$ is the Lande factor), and $\varphi = \eta - \varsigma$. The earlier calculated *real* value of $\langle \Gamma_1 | J_z | \Gamma_2 \rangle$ [11] implies that $\varsigma = 0$. From the time inversion symmetry considerations and a comparison of Eqs.8 and 5 it follows immediately that $W$ is imaginary and $\eta = \pi/2$. Thus, $\varphi = \pi$ and, according to Eq. 7, mixing of the $|\Gamma_1\rangle$ and $|\Gamma_2\rangle$ CF states by an internal magnetic field below $T_N$ does not affect the electron-phonon interaction constant in our case.

The interaction constant $W_0$ and original phonon frequency $\omega_0$ at 300 K were varied to achieve the best agreement with the experimental data. Besides, a linear softening of $\omega_0$ with decreasing the temperature was introduced to account for other than electron-$A_2^1$ phonon interactions, similar to the already discussed case of the $E$ phonon mode near 84 cm$^{-1}$ or the $A_2^1$ phonon mode in SmFe$_3$(BO$_3$)$_4$ (see Inset of Fig. 3). The simulated curves (with $W_0$=14.6 cm$^{-1}$, $\omega_0$=45.5cm$^{-1}$) are plotted in Fig. 3. A good agreement with the experimentally measured temperature dependences of the TO frequencies is evident.

To understand the observed splitting of the IR reststrahlen band, it is necessary to consider a case of a strong (phonon) mode and a weak (electronic) oscillator with the frequency inside the TO-LO splitting of a strong mode. A similar case of a strong and a weak phonons with the oscillator strengths $\Delta \varepsilon_1$ and $\Delta \varepsilon_2$, respectively, $\Delta \varepsilon_2 \ll \Delta \varepsilon_1$, was examined earlier [24] and it was shown that that the LO frequency of a weak phonon becomes *higher* than its TO frequency (an *inverted* phonon). In more detail, for a mentioned pair of non interacting phonons, the upper LO$_+$ frequency shifts upward from the $\omega_{LO1}$ of a strong isolated phonon while the lower LO$_-$ frequency splits from $\omega_{TO2}$ of a weak phonon and shifts downward when increasing $\Delta \varepsilon_2$ (but keeping it small). It is worth adding that this result clearly follows qualitatively from a graphical plot of the $\varepsilon(\omega)$ function of the type (2) for this case. Mathematically, we have the same situation of a weak inverted *electronic* oscillator inside the TO-LO interval of a strong phonon mode. The increase of the electron-phonon interaction with decreasing the temperature leads to a formation of two coupled electron-phonon modes and a redistribution of the oscillator strengths between these two modes. An increasing oscillator strength of a weak mode pushes down the LO$_-$ frequency from the purely electronic frequency $\omega_{12}$, so that a gap in the excitation spectrum develops (see Fig. 3).

To summarize, we combined reflection and ellipsometry measurements in the terahertz region and a theoretical modeling to study electron-phonon coupling in a multiferroic $PrFe_3(BO_3)_4$ single crystal. An isostructural $SmFe_3(BO_3)_4$ single crystal was also investigated, for a comparison. A special feature of $PrFe_3(BO_3)_4$ is that the lowest-frequency $Pr^{3+}$ crystal-field excitation falls into the region between the TO and LO frequencies of a strong lattice phonon mode of the same symmetry as the electronic excitation. Pronounced spectral signatures of the temperature-dependent electron-phonon interaction were observed and explained. In particular, a new effect was demonstrated, namely, a splitting of the reststrahlen reflection band corresponding to a nondegenerate phonon mode. A rather large value of 15 $cm^{-1}$ for the electron-phonon coupling constant was found, which points to an essential role played by the electron-phonon interaction in physics of multiferroics. Our work is also the first observation of an inverted electronic oscillator.

Research supported by the Russian Academy of Sciences under the Programs for Basic Research, the President of Russian Federation under Grants НШ-134.2014.2 and MK-1700.2013.2 (K.N.B.), and the U.S. Department of Energy under Grant No DE-FG02-07ER46382 (experiments at U4-IR beamline NSLS-BNL, T.N.S. and A.A.S.). The National Synchrotron Light Source is operated as a User Facility for the U.S. Department of Energy under Contract No. DE-AC02-98CH10886. M.N.P. thanks B.Z. Malkin for drawing her attention to the work [22] and for helpful discussions.

[1] M. Kaplan, „Cooperative Jahn-Teller Effect: Fundamentals, Applications, Prospects", in *The Jahn-Teller Effect*, Editors: Horst Köppel, David R. Yarkony, Heinz Barentzen (Springer, Berlin, Heidelberg, 2009) Springer Series in Chem. Physics, Vol. 97, pp.653-683 (2009)
[2] K. Ahrens and G. Schaack, Phys. Rev. Lett. **42**, 1089 (1975).
[3] M. Dahl and G. Schaack, Phys. Rev. Lett. **56**, 232 (1986).
[4] A.K. Kupchikov, B.Z. Malkin, A.L. Natadze, and A.I. Ryskin, Fiz. Tverd. Tela **29**, 3335 (1987) [Sov. Phys. Solid State **29**, …..( 1987)l
[5] J. K. Kjems, W. Hayes, and S. H. Smith, Phys. Rev. Lett. **35**, 1488 (1979).
[6] J. Kraus, W. Görlitz, M. Hirsch, R. Roth, and G. Schaack, Z. Phys. B – Condensed Matter **74**, 247 (1989).
[7] A.K. Kupchikov, B.Z. Malkin, A.L. Natadze, and A.I. Ryskin. Spectroscopy of electron-phonon excitations in rare-earth crystals. In *Spectroscopy of crystals* (in Russian), Nauka, Leningrad, 1989, pp. 84-112.
[8] T. V. Brinzari, J. T. Haraldsen, P.Chen, Q.-C. Sun, Y. Kim, L.-c. Tung, A. P. Litvinchuk, J. A. Schlueter, D. Smirnov, J. L. Manson, J. Singleton, and J. L. Musfeldt, Phys. Rev. Lett. **111**, 047202 (2013).
[9] F. Yen, B. Lorenz, Y. Y. Sun, C. W. Chu, L. N. Bezmaternykh, and A. N. Vasiliev, Phys. Rev. B **73**, 054435 (2006); R. P. Chaudhury, F. Yen, B. Lorenz, Y. Y. Sun, L. N. Bezmaternykh, V. L. Temerov, and C. W. Chu, Phys. Rev. B **80**, 104424 (2009); U. Adem, L. Wang, D. Fausti, W. Schottenhamel, P. H. M. van Loosdrecht, A. Vasiliev, L. N. Bezmaternykh, B. Büchner, C. Hess, and R. Klingeler, Phys. Rev. B **82**, 064406 (2010). A. I. Popov, D. I. Plokhov, and A. K. Zvezdin, Phys. Rev. B **87**, 024413 (2013).
[10] D. Fausti, A. Nugroho, P.H.M. Loosdrecht, S.A. Klimin, M.N. Popova, L.N. Bezmaternykh, Phys. Rev. B **74**, 024403 (2006).
[11] M.N. Popova, T.N. Stanislavchuk, B.Z. Malkin, L.N. Bezmaternykh, (a) Phys. Rev. Lett. 102 (2009) 187403; (b) Phys. Rev. B 80 (2009) 195101.
[12] N. V. Kostyuchenko, A. I. Popov, A. K. Zvezdin, Fiz. Tverd. Tela **54**, 1493 (2012) [Phys. Solid State **54**, 1591 (2012)].
[13] S. W. Cheong and M. Mostovoy, Nature Materials, **6**, 13 (2007); J. van den Brink, D. Khomskii, J. Phys.: Condens. Matter **20**, 434217 (2008).
[14] R. Haumont, J. Kreisel, P. Bouvier, and F. Hippert, Phys. Rev. B **73**, 132101 (2006); T. D. Kang, E. Standard, K. H. Ahn, A. A. Sirenko, G. L. Karr, S. Park, Y. J. Choi, M. Ramazanoglu, V. Kiryukhin, and S. W. Cheong, Phys. Rev. B **82**, 014414 (2010); V. S. Bhadram, R. Rajeswaran, A. Sundaresan, and C. Narayana, EPL, **101**, 17008 (2013).
[15] K.N. Boldyrev, T.N. Stanislavchuk, S.A. Klimin, M.N. Popova, L.N. Bezmaternykh, Phys. Lett. A **376**, 2562 (2012).
[16] N.I. Leonyuk and L.I. Leonyuk, Progr. Cryst. Growth Charact. 31, 179 (1995).
[17] A. M. Kadomtseva, Yu. F. Popov, G. P. Vorob'ev, A. A. Mukhin, V. Yu. Ivanov, A. M. Kuz'menko, and L. N. Bezmaternykh, JETP Lett. **87**, 39 (2008).
[18] E.P. Chukalina, M.N. Popova, L.N. Bezmaternykh, I.A. Gudim, Phys. Lett. A **374**, 1790 (2010).
[19] S. A. Klimin, K.N. Boldyrev, to be published.
[20] T. N. Stanislavchuk, T. D. Kang, P. D. Rogers, E. C. Standard, R. Basistyy, A. M. Kotelyanskii, G. Nita, T. Zhou, G. L. Carr, M. Kotelyanskii, and A. A. Sirenko, Rev. Sci. Instr. **84**, 023901 (2013).
[21] A. B. Kuz'menko, D. van der Marel, P. J. M. van Bentum, E. A. Tishchenko, C. Presura, and A. A. Bush, Phys. Rev. B **63**, 094303 (2001).
[22] A.K. Kupchikov, B.Z. Malkin, D.A. Rzaev, and A.I. Ryskin, Fiz. Tverd. Tela **24,** 2373 (1982) [Sov. Phys. Solid State **24,** 1348 (1982)]. It should be emphasized that in the case of crystal excitations the method of local coupled oscillators used, e.g. in [8], is not valid. In particular, it overlooks the dependence of the interaction value on the population difference $n_1$-$n_2$
[23] L.D. Landau & E.M. Lifshitz Quantum Mechanics (Volume 3 of A Course of Theoretical Physics) Pergamon Press 1965.
[24] J. F. Scott and S. P. S. Porto, Phys. Rev. **161**, 903 (1967); F. Gervais, Opt. Comm. **22**, 116 (1977).